\documentclass{bmcart}
\pdfoutput=1

\usepackage{amsthm,amsmath}
\RequirePackage{natbib}
\RequirePackage{hyperref}
\usepackage{subcaption}
\usepackage{graphicx}

\begin{document}

\begin{frontmatter}

\begin{fmbox}
\dochead{Research}

\title{Adding flexibility to clinical trial designs: an example-based guide to the practical use of adaptive designs}

\author[
   addressref={aff1},
   corref={aff1},
   email={t.burnett1@lancaster.ac.uk}
]{\inits{TB}\fnm{Thomas} \snm{Burnett}}
\author[
   addressref={aff1},
   email={p.mozgunov@lancaster.ac.uk}
]{\inits{PM}\fnm{Pavel} \snm{Mozgunov}}
\author[
   addressref={aff2},
   email={pallmannp@cardiff.ac.uk}
]{\inits{PP}\fnm{Philip} \snm{Pallmann}}
\author[
   addressref={aff3},
   email={sofia.villar@mrc-bsu.cam.ac.uk}
]{\inits{SSV}\fnm{Sofia S.} \snm{Villar}}
\author[
   addressref={aff4},
   email={graham.wheeler@ucl.ac.uk}
]{\inits{SSV}\fnm{Graham M.} \snm{Wheeler}}
\author[
   addressref={aff1,aff3},
   email={t.jaki@lancaster.ac.uk}
]{\inits{TJ}\fnm{Thomas} \snm{Jaki}}

\address[id=aff1]{%
  \orgname{Department of Mathematics and Statistics, Lancaster University},
  \street{Fylde College},
  \postcode{LA1 4YF},
  \city{Lancaster},
  \cny{UK}
}
\address[id=aff2]{%
  \orgname{Centre for Trials Research, College of Biomedical \& Life Sciences, Cardiff University},
  \city{Cardiff},
  \cny{UK}
}
\address[id=aff3]{%
  \orgname{MRC Biostatistics Unit, University of Cambridge School of Clinical Medicine,Cambridge Institute of Public Health},
	\street{Forvie Site, Robinson Way, Cambridge Biomedical Campus}
	\postcode{CB2 0SR},
  \city{Cambridge},
  \cny{UK}
}
\address[id=aff4]{%
  \orgname{Cancer Research UK \& UCL Cancer Trials Centre, University College London},
	\street{90 Tottenham Court Road},
	\postcode{W1T 4TJ},
  \city{London},
  \cny{UK}
}

\end{fmbox}

\begin{abstractbox}

\begin{abstract}
Adaptive designs for clinical trials permit alterations to a study in response to accumulating data in order to make trials more flexible, ethical and efficient. These benefits are achieved while preserving the integrity and validity of the trial, through the pre-specification and proper adjustment for the possible alterations during the course of the trial. Despite much research in the statistical literature highlighting the potential advantages of adaptive designs over traditional fixed designs, the uptake of such methods in clinical research has been slow. One major reason for this is that different adaptations to trial designs, as well as their advantages and limitations, remain unfamiliar to large parts of the clinical community. The aim of this paper is to clarify where adaptive designs can be used to address specific questions of scientific interest; we introduce the main features of adaptive designs and commonly used terminology, highlighting their utility and pitfalls, and illustrate their use through case studies of adaptive trials ranging from early-phase dose escalation to confirmatory Phase~III studies. 
\end{abstract}

\begin{keyword}
\kwd{novel designs}
\kwd{innovative trials}
\kwd{efficient methods}
\kwd{enrichment designs}
\kwd{multi-arm multi-stage platform trials}
\end{keyword}

\end{abstractbox}

\end{frontmatter}

\section{What are adaptive designs?}

\begin{figure}[!ht]
\begin{subfigure}{\textwidth}
\makebox[\textwidth][c]{\includegraphics[scale=.8]{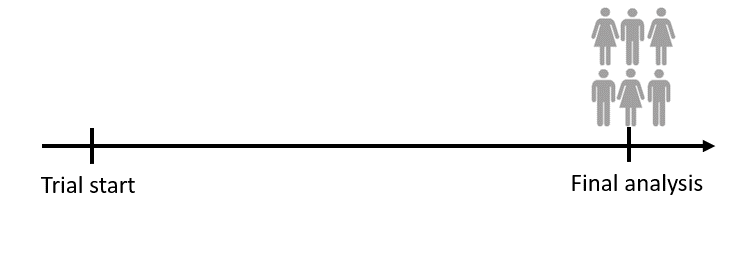}}
\caption{Fixed design analysis plan.}
\label{fig:fva1}
\end{subfigure}
\begin{subfigure}{\textwidth}
\makebox[\textwidth][c]{\includegraphics[scale=.8]{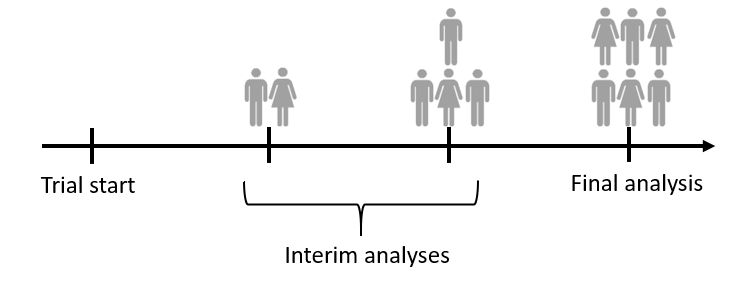}}
\caption{Adaptive design analysis plan.}
\label{fig:fva2}
\end{subfigure}
\caption{A comparison of fixed and adaptive designs analysis plans.}
\end{figure}

In a traditional clinical trial the design is fixed in advance and the study then carried out as planned with the data being analysed after completion \cite{friedman2010fundamentals} as illustrated in Figure~\ref{fig:fva1}. Such a trial design is routine in clinical research, but clearly inflexible whenever modifications of the planned trial conduct become desirable or necessary unless a protocol amendment is made. In contrast, adaptive designs pre-plan possible modifications on the basis of the data accumulating over the course of the trial. Aspects of a trial that can be modified include: the sample size, the number of treatment arms, or the allocation ratio to different arms. With these modifications pre-planned as part of an adaptive design, these changes can be made without requiring a protocol amendment. Typically these modifications are made based on data available within the study. It is adaptations of this nature that we consider here, while we do not consider options such as stopping early due to failure to meet operational criteria or excessive safety events; although adaptive designs for some of these do also exist \citep{hampson2018framework}. Adaptations to a trial in progress are made using pre-planned interim analyses \cite{pallmann2018adaptive}. Figure~\ref{fig:fva2} demonstrates how interim analyses are a part of the pre-planned design demonstrating the key difference between adaptive and fixed (non-adaptive) designs.\\

The defining characteristic of all adaptive designs we discuss in this work is the use of pre-planned interim analyses to decide if a pre-planned modification of the ongoing trial should be undertaken, without undermining its integrity and validity \cite{chow2005statistical}.  This is crucial: the integrity of a trial relates to the trial data and a guarantee that the data have not been used in such a way as to substantially alter the result, while the validity of the results requires that the study answers the original research questions appropriately. In an adaptive design, data are looked at more often than in a traditional study, thus the trial requires further procedures to ensure that data is collected, analysed, and stored in an appropriate manner at every stage of the trial. To ensure the validity of an adaptive design, specialised statistical methodology is required for inference; for example, to provide correct p-values, unbiased estimates, and confidence intervals for the treatment comparison(s). By their nature adaptive designs are more complex than their fixed sampling counterparts and this should be reflected in the reporting \cite{dimairo2018development}.\\

The possibility to make changes during the trial is not a virtue in itself but rather a gateway to more efficient and ethical trials in comparison to traditional fixed designs: futile treatment arms may be dropped sooner, more patients may be randomised to a superior treatment, fewer patients may be required overall, treatment effects may be estimated with greater precision, a definitive conclusion may be reached earlier, etc. Adaptive designs can aid in these aspects across all phases of clinical development \cite{pallmann2018adaptive} (see Table~\ref{tab:glos} for examples of adaptive designs) and therefore lead to more efficient and accurate decision-making throughout the development process for novel treatments and interventions. Furthermore, seamless designs make use of adaptive methods that allow for smooth and speedy transitions between trials of different phases \cite{zohar2007recent,sverdlov2014novel,stallard2011seamless,maca2006adaptive}. \\

How the adaptation decisions are made is a crucial part of the design of an adaptive trial. It is often possible to extend beyond classical statistical techniques and consider the use of a Bayesian methodology in the decision making process. Bayesian methods allow the incorporation of prior information (such as the current state of knowledge based on a review of the literature or expert opinion) into the design and analysis of a trial, provide more intuitive interpretation, and offer many other benefits \cite{jansen2017bayesian}. Thus they fit naturally within the setting of adaptive designs.\\

Adaptive design methodology has been around for more than 25 years \cite{bauer2016twenty}, with some methods such as group sequential designs being even older \cite{pocock1977group}. Despite the many clear benefits, adaptive designs are still far from established as typical practice \cite{le2009dose,jaki2013uptake,chevret2012bayesian}. Many reasons for this have been identified \cite{jaki2013uptake,chevret2012bayesian,dimairo2015cross,dimairo2015missing} the main of which include: lack of expertise and experience (specific to the use of a suitable adaptive design); lack of software to assist in design and analysis; longer time required for planning and analysis of the trial; inadequate funding structure to account for uncertainty about how the trial will run; and the fact that chief investigators may insist on using a particular method they are comfortable with.  We believe that, besides their knowledge of previous studies in the area or their own past experience with traditional designs, the main reason why investigators are not inclined to adopt adaptive designs is a lack of clarity among clinicians and trialists about what these are and what they can (and cannot) accomplish. Ambiguous terminology with vague definitions adds to any confusion \cite{dragalin2006adaptive}. To address this ambiguity, we provide a glossary of the most common types of adaptive design in Table~\ref{tab:glos}.\\

To help overcome some of the uncertainties and demonstrate when adaptive designs can be useful we focus on four key questions of scientific interest when developing and testing novel treatments: `What is a safe dose?'; `Which is the best treatment among multiple options?'; `Which patients will benefit?'; `Does the treatment work?'. For each of these questions we shall briefly review several important adaptive designs, and outline their advantages and disadvantages. We shall illustrate methods that can be used from early to late phase clinical development, and provide real world examples of the designs. Finally, we discuss when the use of an adaptive design is the best course of action, potentially worthwhile, or not advisable. 

\section{When to use an adaptive design}
We identify four key clinical questions that must be answered when developing a new treatment across the Phase~I, II and III trials (for drug development, terminology differs for surgical or behavioral interventions). We discuss adaptive designs appropriate to address each of these questions during the clinical development process. We shall detail the purpose of each method, outlining both advantages and disadvantages of their implementation and provide examples of their use in practice.\\

\subsection{What is a safe dose?}
\label{sec:1}

Phase I trials of new drugs are conducted to assess the safety of a treatment, the aim being to establish the safety profile across a range of available doses. The data collected are then used to select a dose for further testing in larger Phase~II trials, in which the primary objective is to establish the potential efficacy of the treatment. In many therapeutic areas, the goal of a Phase~I trial is to identify the maximum tolerated dose (MTD), that is the highest dose that controls the risk of unacceptable side effects \cite{national2014national} and hence is deemed safe. In practice, one seeks to identify the dose at which the probability of a dose-limiting toxicity (DLT) is equal to some pre-specified target level, usually around 20-33$\%$. In oncology, this is usually done by treating consenting patients sequentially at increasing doses until too high a proportion of unacceptable side effects are observed. Several approaches have been proposed for conducting Phase~I dose-escalation studies; we provide a summary of the most common of these approaches here and show examples of the corresponding adaptive designs being used in practice.\\

\subsubsection{3+3 design}
\begin{figure}[!ht]
\makebox[\textwidth][c]{\includegraphics[scale=.8]{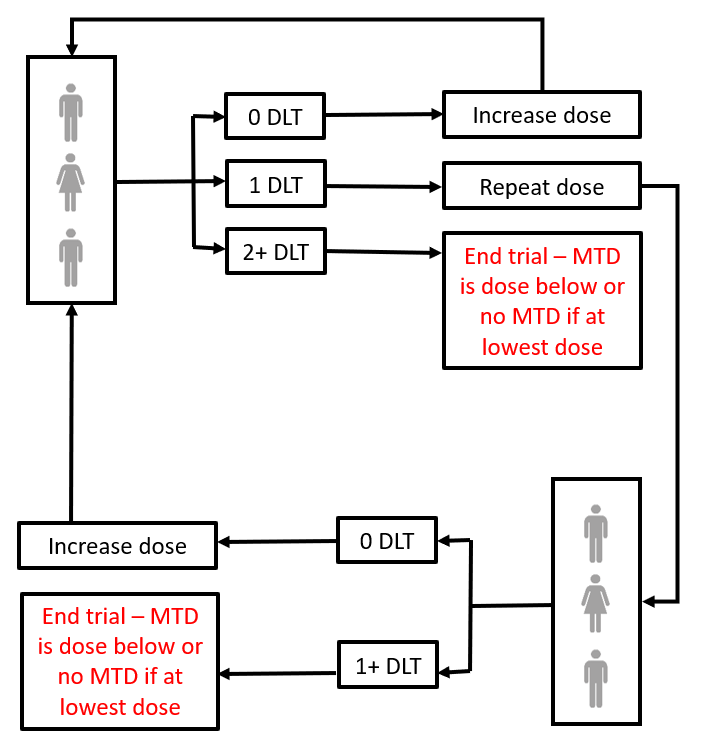}}
\caption{Decision making in a 3+3 design based on the current cohort.}
\label{fig:3+3}
\end{figure}

The most commonly used method for conducting dose-escalation studies is the 3+3 design \cite{carter1973study,storer1989design}. It is a simple, rule-based approach under which patients are dosed in cohorts of three. Based on the number of DLTs observed in the current cohort of patients, recommendations are made to dose the next three patients at either the next escalating dose or the current dose. Extending upon the original 3+3 design a third option of reducing the dose below the current level may be included. Upon observing excessive toxicity at a dose level (say DLTs in more than 2 in 6 patients), the trial is terminated and the dose level below is considered to be the MTD. A simple illustration of this rule is shown in Figure~\ref{fig:3+3}. Rule based methods such as 3+3 designs are adaptive, in the sense that the outcomes of the current cohort determine the dose for the next cohort.\\

\underline{Example:} Park et al. \cite{park2005phase} performed a Phase I dose-escalating study of docetaxel in combination with 5-day continuous infusion of 5-fluorouracil (5-FU) in patients with advanced gastric cancer. The study used a 3+3 design to find the MTD. The treatment consisted of docetaxel 75 mg/m$^2$ on day 1 in a 1 h infusion followed by 5-FU in continuous infusion from day 1 to day 5, according to the escalating dose levels. The starting dose of 5-FU was 250 mg/m$^2$/day for 5 days. In the absence of any DLTs (defined as as febrile neutropenia and/or grade 3/4 toxicity of any other kind apart from alopecia), dose escalation in additional cohorts continued, increasing the dose by 250 mg/m$^2$/day for each increment.\\

Of three patients enrolled on dose level 1 (5-FU 250 mg/m$^2$/day for 5 days), none experienced DLT. On dose level 2 (5-FU 500mg/m$^2$/day for 5 days), one patient developed grade 3 stomatitis and fatigue. Three additional patients were enrolled at this dose level, none of whom experienced any DLT. Thus dose escalation proceeded to dose level 3 (5-FU 750 mg/m$^2$/day for 5 days). One patient developed grade 3 stomatitis and another patient had febrile neutropenia. Because two of the three patients had DLTs dose escalation was stopped at dose level 3. Dose level 2 was therefore the recommended regimen with docetaxel 75 mg/m$^2$ on day 1 and 5-FU 500 mg/m$^2$/day in a 5-day continuous infusion.\\

\underline{Advantages:} The key advantage of the 3+3 design is that it does not require any time to design. Web applications \cite{wheeler2016aplusb} are available to understand the performance of such designs. In addition this method is well-known to clinicians, often leading to its use being well motivated within the trial team.\\

\underline{Disadvantages:}
The major disadvantages of the 3+3 design will become clear as we draw comparisons to the methods that follow. In particular we note that the MTD is not explicitly defined; this means that the most likely dose to be chosen as the MTD can have a probability of DLT far from the assumed target and can be highly variable \cite{kang2001expected,kang2002investigation,he2006model}.\\

Rule based dose escalation methods such as 3+3 designs are seriously flawed, which runs afoul of the part of our definition of an adaptive design that demands integrity and validity. Thus this method being well-known to clinicians, possibly allowing them to avoid collaboration with a statistician can also present a serious problem. \\

\subsubsection{Continual Reassessment Method (CRM)}
\begin{figure}[!ht]
\makebox[\textwidth][c]{\includegraphics[scale=.8]{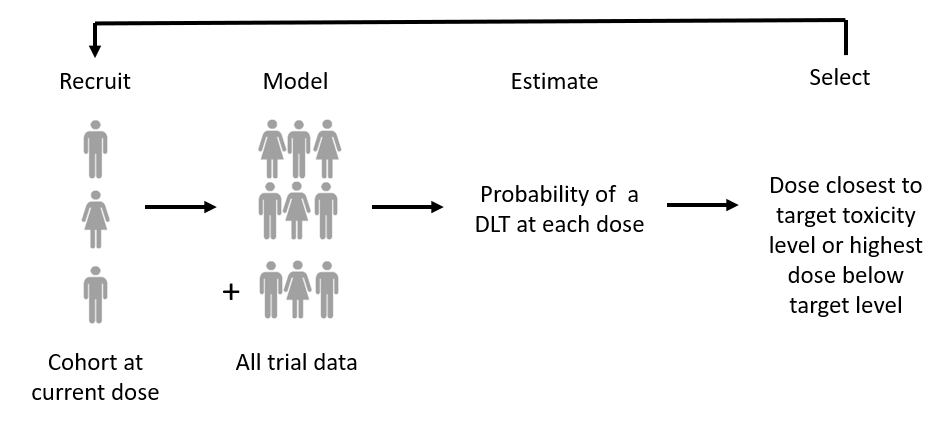}}
\caption{Steps within each iteration of a CRM design.}
\label{fig:CRM}
\end{figure}

The Continual Reassessment Method (CRM) \cite{o1990continual, o1996continual} uses a simple function to model the relationship between dose and the risk of a patient experiencing a DLT. Rather than counting how many people have had DLTs at each dose to determine whether to dose-escalate or not, all available trial data (and possibly relevant prior information on the safety profile of the treatment) are used to estimate the probability of a DLT at each dose level. For a target toxicity level of interest, the dose for the next patient or cohort is either that with an estimated probability of DLT closest to the target level, or the highest available dose below the target level. This process is iterated for each new cohort of patients, as illustrated in Figure~\ref{fig:CRM}.\\

\underline{Example:} Paoletti et al. \cite{paoletti2006using} describe the design, conduct and analysis of a multicentre Phase~I trial to find the MTD (defined as the dose with probability of a DLT closest to $20\%$) of rViscumin in patients with solid tumours. A DLT was defined as any haematological grade 4 or non-haematological grade 3 or grade 4 adverse event as defined by the National Cancer Institutes Common Terminology Criteria for Adverse Events (NCI CTCAE) Version 2, with the exclusion of nausea, vomiting, or rapidly controllable fever. The starting dose of the trial was 10 ng/kg, with fixed dose levels for further exploration of 20, 40, 100, 200, 400, 800 ng/kg; if no adverse events of grade 2 or higher were observed after escalation to 800 ng/kg, additional doses would be added in increments of 800 ng/kg (i.e. 1600, 2400, etc ng/kg).\\

The design for dose escalation used a two-stage CRM \cite{o1990continual}. First, one patient was assigned to the starting dose of 10 ng/kg and, if adverse events were absent or grade 1, a new patient was given the next highest dose; if a non-DLT adverse event of grade 2 or higher was observed, a further two patients would be given the same dose. Escalation continued in this manner until the first DLT was observed, at which point the model-based design took over. A one-parameter model \cite{o1996continual} was fitted to the data, and the dose with an estimated probability of DLT closest to 20$\%$ was recommended for the next patient; subject to the constraint that no untested dose level is skipped. The trial was stopped when the probability of the next five patients being given the same dose was at least 90$\%$ (i.e. the trial would be unlikely to gain further information that would affect dose allocation).\\

The first DLT was observed in the 11th patient who was given 4000 ng/kg; at which point the CRM part of the design took over. In total 37 patients were recruited the trial before it was terminated under the aforementioned rule, and the MTD declared as 5600 ng/kg, with an estimated DLT probability of 16$\%$; the estimated probability of a DLT at the next highest dose (6400 ng/kg) was 31$\%$.  It is worth noting that the first DLT was recoded during the trial to a non-DLT, without this a lower MTD would have been selected \cite{paoletti2006using}; this illustrates one of the benefits of a model-based approach; deviations from the planned course of the trial are handled without compromising the validity of the design, we can simply re-estimate our DLT risks at each dose and carry on as normal \cite{wheeler2019design}.\\

One DLT was observed at dose level 4800 ng/kg (patient 26), and two DLTs were observed at dose level 6400 ng/kg (patients 35 and 37). The description of this trial by Paoletti et al. \cite{paoletti2006using} is itself a tutorial of the practical considerations for designing a Phase~I trial using the CRM. The authors describe how the statistical work of the trial helped to inform study clinicians and the Trial Steering Committee, with whom any final decisions rest. For example, a decision was made to dose another patient at 3200 ng/kg rather than escalate to 4000 ng/kg as per the design in order to gather more PK data at this level. Furthermore 10 extremely tolerable (but presumably inefficacious) dose levels were cleared quickly and with far fewer patients than the 3+3 design would require.\\

\underline{Advantages:} Conceptually, the CRM is a far wiser approach (and more efficient/economic) than the 3+3 design because it uses all available trial data to make decisions, rather than solely the data from the last cohort \cite{ratain1993statistical,o2006experimental}.  Also, the CRM targets a pre-specified toxicity level, whereas the 3+3 design provides an MTD that is not defined explicitly with respect to a risk of a DLT. Numerous comparative simulation studies have shown the CRM to supersede the 3+3 design and other competing statistical approaches by dosing more patients in trial at or near the correct MTD and also by selecting the correct dose as the MTD at the end of the trial more often \cite{o1999two,thall2003dose,iasonos2008comprehensive,onar2009continual,onar2010simulation}, which in turn can result in higher probability of success in subsequent Phase~II and Phase~III clinical trials \cite{conaway2019impact}. \\

Furthermore, the CRM can easily be adapted to include more informative endpoints such as: multiple graded toxicities to incorporate severity of side effects \cite{lee2010continual,iasonos2011incorporating,van2012dose}; combinations of safety and  efficacy outcomes \cite{braun2002bivariate,zohar2006optimal,zohar2006identifying,zhong2012trivariate,yeung2015bayesian,yeung2017bayesian}; time-to-event outcomes to distinguish between toxicity events occurring sooner or later \cite{cheung2000sequential,braun2006generalizing}; or even developed to escalate multiple treatments at once \cite{kramar1999continual,wang2005two,yuan2008sequential,wages2011dose,harrington2013adaptive,riviere2014bayesian,riviere2015competing,wages2016statistical}. Regulatory authorities are also recognising that novel adaptive designs using statistical models are of great importance, and actively encourage sensible usage of them in Phase~I trials \cite{guidance2019adaptive,nie2016rendering}.\\

\underline{Disadvantages:} The main disadvantage of the CRM design is that more time and effort is required at the design stage of the trial to assess how it is expected to perform. This requires close collaboration between the clinical team and a suitably trained statistician who is able to guide this optimisation process; although this opportunity to consider the study more carefully before it begins can only be a good thing for the study on the whole. The clinical team may still see the CRM as a ``black box''; to resolve this concern, Dose Transition Pathways provide a tool for visualisation of the CRM escalation/de-escalation decisions \cite{yap2017dose}, which can assist in the communication of the design. Several computer programs are available (see MD Anderson Cancer Center software library \cite{MDAC2020}, Vanderbilt University, and packages within R \cite{bove2019model,sweeting2013bcrm}) for conducting simulation studies, some of which can offer comparisons to other popular dose escalation designs including the 3+3 design \cite{sweeting2013bcrm} and tutorial papers are available offering further guidance \cite{wheeler2019design}. Web based solutions are on offer for both the CRM and conceptual equivalents \cite{wages2018web,chen2017web,pallmann2019designing}.

\subsubsection{Escalation with Overdose Control}

The Escalation with Overdose Control (EWOC) approach \cite{babb1998cancer} to Phase~I trials is similar to the CRM in that all patient data in a trial is used to make dose-escalation decisions and a target toxicity level is used to choose which dose level the next patient or cohort should receive, with a few key differences. Firstly, rather than choosing the next patient's dose as the estimated DLT probability closest to the target level, the EWOC approach assigns the next patient using a skewed allocation criterion to account for the fact that the overdosing of patients is much more undesirable compared to the underdosing. This results in a much more conservative patient allocation approach, with fewer patients being exposed to possible overdosing and experiencing a DLT compared to the CRM, while still benefiting from the model attempting to allocate the patients near the MTD \cite{wheeler2017toxicity,tighiouart2010dose}. Secondly, the same statistical model is re-expressed in a way that allows focus on the clinically relevant parameters, the MTD, and the probability of a DLT at the lowest dose. This means that prior information about the treatment being investigated can easily be incorporated and one can visualise how the distribution of the MTD changes over the course of the trial.\\

\underline{Example:} Nishio et al. \cite{nishio2015phase} conducted a Phase~I dose-escalation study of ceritinib in patients with advanced anaplastic lymphoma kinase-rearranged, non-small-cell lung cancer or other tumours. A Bayesian EWOC approach was used to govern dose-escalation, allocating the next patient to the largest dose with an estimated probability of less than $25\%$ that the risk of a DLT exceeds $33\%$. In total, 19 patients were recruited to the trial: three patients received doses of 300 mg, 6 patients received doses of 450 mg, four patients received doses of 600 mg and six patients received doses of 750 mg. Two patients experienced DLTs, one at 600 mg and the other at 750 mg. At the end of the trial, the MTD was chosen as 750 mg; the largest dose at which the estimated probability of the risk of a DLT exceeding $33\%$ was less than the target $25\%$ (the probability was $7.3\%$ for the chosen dose). Although the aim of the trial was not to evaluate efficacy of ceritinib in this population, 10 patients achieved partial responses to their cancers.\\

\underline{Advantages:} The EWOC approach offers a more cautious dose-escalation design that reduces the chance of patients being treated at excessively toxic doses \cite{babb1998cancer}. However, a slower dose-escalation approach may increase the number of patients treated at sub-therapeutic doses; although this also means that fewer patients are treated at doses close to the MTD. Similar to the CRM, the EWOC approach has also been adapted to be used in trials with more complex outcomes, such as time-to-event data \cite{tighiouart2014escalation}, and for combinations of treatments \cite{shi2013escalation,tighiouart2014dose}. Furthermore, the escalation control threshold can be altered depending on the trial context, and may change during the conduct of the trial \cite{tighiouart2010dose,wheeler2017toxicity}; this offers a conservative dose-escalation schema at the start of the trial when there is little data available, but as more data are accrued, dose-escalation gradually becomes less conservative, and the MTD can be targeted more quickly than with the standard EWOC approach \cite{wheeler2017toxicity,tighiouart2010dose,mozgunov2019improving}.\\

\underline{Disadvantages:} Similar to implementation of the CRM, care is required when designing trials using the EWOC approach. For example, choice of the MTD estimator needs to be considered; several trials use the same criterion as that by Nishio et al. \cite{nishio2015phase}, i.e. the MTD is the dose that would be given to a new patient had they entered a trial, whereas others have used the posterior median of the MTD distribution. The implications of each choice need to be considered well in advance \cite{berry2010bayesian}. Furthermore, if the investigators plan to relax the escalation control mechanism (that is, decrease how cautious escalation is as the trial continues), as has been done in practice before \cite{babb2001patient,cheng2004individualized,tighiouart2010dose}, the implications of this decision need to be considered. The EWOC approach may recommend to escalate the dose even when the most recently evaluated patient experienced a DLT \cite{wheeler2018incoherent}.

\subsubsection{Summary}

Despite the 3+3 designs frequent use in Phase~I clinical trials over the last 30 years \cite{rogatko2007translation,le2009dose,le2012efficiency,rivoirard2016thirty}, there is overwhelming consensus among statisticians and methodologists that it is sub-optimal, and more efficient designs for identifying the MTD should be used \cite{paoletti2015statistical,conaway2019impact,jaki2013principles}.  Many alternative designs propose the use of statistical models, such as the two alternatives we have presented here; both of which have superior operating characteristics over the 3+3 design.\\

The two model-based approaches discussed above serve as the main frameworks for other proposed model-based approaches, which are designed for trials with novel drug combinations, endpoints that use time-to-event data and/or efficacy outcomes, or information about the severity of observed toxicities. These designs have slowly found their way into clinical practice in recent years, primarily in oncology for cytotoxic treatments. However, these approaches can be used for novel molecularly targeted anti-cancer therapies \cite{mandrekar2007adaptive}, and in other disease areas altogether: O’Quigley et al. \cite{o2001dose} proposed CRM-type designs for use in dose-finding studies for anti-retroviral drugs to treat Human Immunodeficiency Virus (HIV); Lu et al. \cite{lu2016phase} conducted a dose-escalation study of quercetin in patients with Hepatitis C; Whitehead et al. \cite{whitehead2006bayesian} proposed a model-based design for trials in healthy volunteers; and Lyden et al.\cite{lyden2019randomized} used a CRM design in the RHAPSODY trial in stroke patients.\\

Whilst there is no “one size fits all” approach for conducting adaptive dose-escalation studies, there is overwhelming evidence that model-based designs are far better than the standard rule-based designs, such as the 3+3 design. Model-based designs are on the whole more efficient in their use of data, less likely to dose patients at subtherapeutic doses, more likely to recommend the correct MTD at the end of the trial, and provide an MTD estimate that directly relates to a specified target level of toxicity. We have discussed two approaches here for brevity, though many other alternatives have also been proposed, including designs based on optimal design theory \cite{haines2003bayesian,azriel2014optimal,haines2014construction,liu2015bayesian} and model-free designs \cite{gasparini2000curve,mander2015product,yuan2016bayesian,mozgunov2019information}, without the shortcomings of common rule-based designs like 3+3. The increasing usage of model-based designs in practice, as well as their acknowledgement in regulatory guidance and the provision of guideline documents \cite{lorusso2010overview}, formal courses and computer software is indicative of the changing tide of clinical practice for Phase~I trials.\\

\subsection{Which is the best treatment among multiple options?}
\label{sec:2}

After establishing the safety of a treatment the next question is naturally ``how does it perform?''. In this section, we consider randomised clinical trials that aim to select the best treatment arm among multiple experimental treatment arms (where these can be different treatments, doses of the same treatment or combinations of these two options). The methods we explore are typically considered for use in Phase~II of the development process, where we wish to select a treatment or dose for further study. A common theme is to compare multiple arms against a common control. We explore methods that seek to remove less beneficial treatments from the trial quickly, giving patients a better chance of receiving an efficacious treatment. In addition model-based approaches allow modelling of the dose-response relationship in order to provide a deeper understanding of this in an efficient way. This is useful in trials where the different arms correspond to a series of exposure levels (such as doses of a drug, duration or intensity of radiotherapy, or number of therapy sessions).\\

\subsubsection{Multi-Arm Multi-Stage (MAMS)}
\label{sec:MAMS}

\begin{figure}[!ht]
\makebox[\textwidth][c]{\includegraphics[scale=.8]{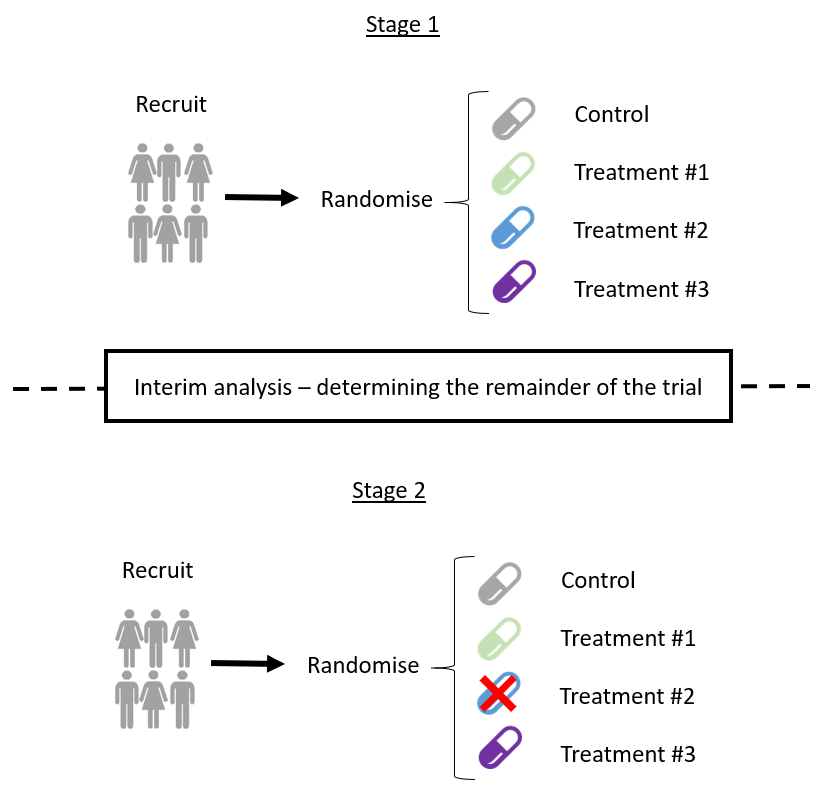}}
\caption{A two-stage four-arm MAMS design}
\label{fig:MAMS}
\end{figure}

Multi-Arm Multi-Stage (MAMS) \cite{jaki2015multi} trials allow the simultaneous testing of multiple experimental treatment arms with a single common control. They are conducted over multiple stages to allow for the early stopping of recruitment for either efficacy or futility. For example, if an experimental treatment is found to be performing poorly it may be dropped for futility at a pre-planned interim analysis (if all experimental arms are dropped, the trial is stopped for futility). Alternatively, the trial may end early when a treatment is shown to be sufficiently efficacious. We cover group sequential designs in more detail in Section~\ref{sec:gs} but MAMS designs apply similar methodology while comparing multiple experimental treatments simultaneously. A simplex schematic of how a two-stage four-arm trial using MAMS design can progress is given in Figure~\ref{fig:MAMS}.\\

MAMS trials are often designed using a pre-planned set of adaptation rules to find the best treatment to carry forward for further study \cite{stallard2003sequential} or carry forward all promising treatments \cite{magirr2012generalized}. Alternatively more flexible testing methods \cite{bauer1999combining, bretz2006confirmatory, schmidli2006confirmatory} allow methodological freedom as to how adaptation decisions are made. Open source software is available to assist in the design and analysis of MAMS trials in the form of the ``MAMS'' package for R \cite{jaki2019r}. Alternatively in STATA there are several modules available such as ``nStage'' \cite{royston2019nstage}, ``nStagebin'' \cite{bratton2014nstagebin}, and ``DESMA'' \cite{grayling2019desma}.\\

\underline{Example:} The TAILoR trial \cite{pushpakom2015telmisartan} was a Phase~II, multicentre, randomised, open-labelled, dose-ranging trial of telmisartan using a two-stage MAMS design. The trial planned recruitment of up to 336 HIV-positive individuals over a 48-week period, with a single interim analysis planned after 168 patients had completed 24 weeks on either an intervention or control treatment. Patients were randomised with equal probability to one of four groups: no treatment (control), 20mg telmisartan daily, 40mg telmisartan daily, or 80mg telmisartan daily.\\

At the interim analysis there were three possible outcomes based on assessment of change in HOMA-IR index from baseline to 24 weeks: if one telmisartan dose was substantially more effective than control, the study would stop and that dose would be recommended for further study; if all telmisartan doses were less effective than control, the study would stop with no dose recommended for further study; if one or more doses were better than control but none met the first criterion, the study would continue and patients would have been randomised between these remaining dose(s) and control. If the trial entered the second stage then a final analysis would be conducted with two possible outcomes based on the primary endpoint: either the best dose is significantly more effective than the control in which case it is recommended for further study; or no dose is significantly better than control in which case no dose is recommended.\\

A total of 377 patients were recruited \cite{TAILORres} (note this difference in sample size was due to higher than expected drop-out). In stage one, 48, 49, 47, and 45 patients were randomised to control and 20, 40, and 80 mg telmisartan, respectively. Following the interim analysis only 80 mg telmisartan was taken forward into stage two. At the end of stage two 105 patients had been recruited to control and 106 to the 80-mg arm (in total), there was no difference in HOMA-IR (estimated effect, 0.007; SE, 0.106) at 24 weeks between the telmisartan (80 mg) and control arm.\\

\underline{Advantages:} MAMS designs are useful when there are multiple promising treatments and there is no strong belief that one treatment will be more beneficial compared to the others. The use of a shared control group considerably reduces the number of patients that need to be recruited compared to separate RCTs testing each treatment. Other advantages are: treatments that provide no benefit to patients are dropped from the trial leading to a potentially smaller sample size; patients have a higher chance of receiving an experimental treatment compared to a 2-arm trial, which may improve recruitment to the study \cite{dumville2006use,meurer2012adaptive}; they require less administrative and logistical effort than setting up separate trials and thus can substantially speed up the development process \cite{parmar2008speeding}.\\

\underline{Disadvantages:} MAMS trials require an outcome measure that allows a timely decision about the worth of each treatment. Consequently either the primary endpoint needs to be relatively quickly observed (in comparison to patient accrual) or an intermediate measure that is strongly associated with the primary endpoint is required for interim decision making. MAMS designs will generally require a larger potential maximum sample size than a corresponding multi-arm fixed design (although still smaller than several separate trials due to the use of a common control arm). The MAMS approach has a variable sample size depending on which decisions are made through the trial, which will make planning more cumbersome, although the possible pathways are pre-defined; this is more variable than is typical of even other adaptive designs because decisions relate to each treatment individually.\\

\subsubsection{Drop The Loser (DTL)}

Drop The Loser (DTL) designs \cite{thall1989two,sampson2005drop} are closely related to MAMS designs \cite{wason2017multi} in that they aim to compare several experimental treatments to a common control over multiple stages.  The key difference between the two methods is that in a DTL design it is pre-determined how many arms will be dropped after each stage of the trial. As the name suggests the worst performing experimental treatments are dropped at interim analysis, leaving only one treatment to compare to control at the final analysis.\\

\underline{Example:} The ELEFANT trial \cite{zadori2019early} is a randomised controlled, multicentre, three-armed trial testing the concept that early elimination of triglycerides and toxic free fatty acids from the blood is beneficial in  HyperTriGlyceridemia-induced Acute Pancreatitis (HTG-AP). This study will be conducted using a two-stage DTL design, dropping the inferior experimental treatment arm at the interim analysis. Patients with HTG-AP are randomised with equal probability into three groups: patients who undergo plasmapheresis and receive aggressive fluid resuscitation; patients who receive insulin and heparin treatment with aggressive fluid resuscitation; and patients with aggressive fluid resuscitation only (the control). The target sample size is 495 in order to detect a $66\%$ relative risk reduction, using a $10\%$ dropout rate with $80\%$ power at $5\%$ significance level. The the study began in February 2019 and is expected to finish December 2024.\\

\underline{Advantages:} As with MAMS designs the key advantage is that the use of the shared control group considerably reduces the number of patients required. Of further practical benefit is that there is a guarantee that a pre-specified number of arms will be dropped during the course of the trial, meaning that the required sample size is known before commencement of the trial \cite{wason2017multi}. At the conclusion of the trial only one experimental treatment remains to be compared against the control making for a clear interpretation of results.\\

\underline{Disadvantages:} The choice to only continue the most efficacious treatments may not always be the best choice; consider for example an interim analysis where a dropped treatment has demonstrated an almost equivalent positive effect to a treatment that continues in the trial, it is possible that a suitable treatment has been dropped by chance. In addition a similar operational complexity to MAMS designs is observed here as it is unknown which treatments will be carried through the trial (although as noted previously the exact required sample size is known which it is not for a MAMS design).

\subsubsection{Response-Adaptive Randomisation (RAR)}
\label{sec:RAR}

\begin{figure}[!ht]
\makebox[\textwidth][c]{\includegraphics[scale=.8]{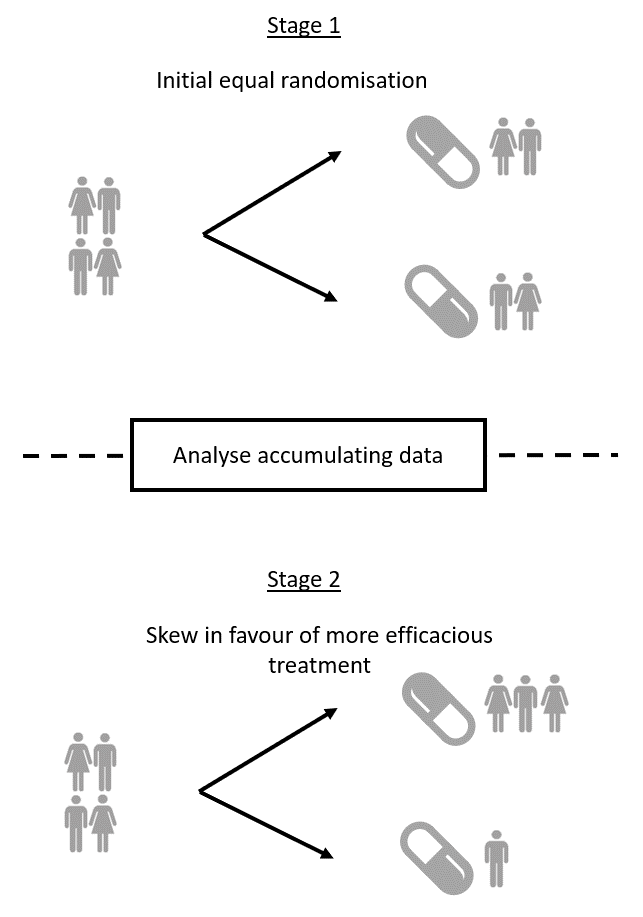}}
\caption{A description of a RAR procedure.}
\label{fig:RAR}
\end{figure}

At the beginning of the trial, usually not much is known about the efficacy of the experimental treatment arms; hence, equal randomisation is often sensible under a clinical equipoise principle. However, as data accumulates it becomes challenging from an ethical standpoint to randomise patients to a treatment arm that accumulating trial data suggest may be inferior, especially when the disease under study is rare or life-threatening. One way to resolve this dilemma is to align the randomisation probabilities with the observed efficacy of the different arms, which is exactly what Response-Adaptive Randomisation (RAR) does (Figure~\ref{fig:RAR}).\\

The idea of RAR dates back to 1933 \cite{thompson1933likelihood} and since then several methods to align randomisation probabilities and observed evidence of efficacy have been proposed in the specialised literature \cite{berry1995adaptive,hu2006theory}. Thompson \cite{thompson1933likelihood} proposed to randomise patients to arms with a probability that is proportional to the probability of an arm being the best arm. Regardless of how these probabilities are defined and applied for a specific adaptation rule, they can be also used to define further adaptations to the trial depending on the values they assume. For example, if the allocation probability goes below or rises above a certain value, arms can be dropped for futility or selected in a similar way to a MAMS study \cite{wason2014comparison,lin2017comparison}. Free software is available from the MD Anderson website \cite{MDAC2020}. The R package ``bandit'' offers an alternative to the implementation of such designs \cite{lotze2015bandit}.\\

\underline{Example:} A prospective, randomised study reported by Giles et al. \cite{giles2003adaptive} was conducted in patients aged 50 years or older with untreated, adverse karyotype, acute myeloid leukemia to assess three troxacitabine-based regimes: idarubicin and cytarabine (the control arm); troxacitabine and cytarabine; and troxacitabine and idarubicin. The trial used a Bayesian RAR design along the lines proposed by Thompson \cite{thompson1933likelihood}. Thirty-four patients were recruited and randomised to one of the three arms.  Initially an equal chance for randomisation to each arm was ensured, but treatment arms with a higher success rate, defined as the proportion of patients having complete remission within 49 days of starting treatment, would receive a greater proportion of patients. The randomisation probabilities were updated after every patient. The design would drop arms if their assignment probabilities became too low or promote them to Phase~III if their assignment probability was high enough. The probability of a patient being randomised to the control arm was fixed until the first experimental arm was dropped. This occurred when the randomisation probability for the dropped experimental arm was 0.072 \cite{grieve2017response}.\\

Of the thirty-four patients recruited 18 were randomised to idarubicin and cytarabine, randomisation to troxacitabine and idarubicin stopped after five patients and randomisation to troxacitabine and cytarabine stopped after 11 patients. Success rates were $55\%$ (10 of 18 patients) with idarubicin and cytarabine, $27\%$ (three of 11 patients) with troxacitabine and cytarabine, and $0\%$ (zero of five patients) with troxacitabine and idarubicin, while survival was comparable with all three regimens.\\

\underline{Advantages:} RAR can increase the overall proportion of patients enrolled in the trial who benefit from the treatment they receive while controlling type~I (the probability of falsely rejecting the null hypothesis when it is true, a false positive) and type~II error rates (the probability of failing to reject the null hypothesis when it is false, a false negative) \cite{villar2018response,gutjahr2011familywise,robertson2019familywise}. This mitigates potential ethical conflicts \cite{london2018learning} that can arise during a trial when equipoise is broken by accumulating evidence and makes the trial more appealing to patients \cite{meurer2012adaptive}. Therefore, the RAR design may improve the recruitment rates \cite{tehranisa2014can}. The main motivation for RAR designs is to ensure that more trial participants receive the best treatments, it is possible to use such response adaptive methods to optimise other characteristics (such as power) of the trial \cite{hu2003optimality}. In a multi-armed context RAR can shorten the time of development and more efficiently identify responding patient populations \cite{berry2010adaptive}.\\

\underline{Disadvantages:} RAR designs have been criticized for a number of reasons \cite{proschan2020} although many of the  raised concerns can be addressed. Logistics of trial conduct is a noticeable obstacle in RAR due to the constantly changing randomisation \cite{korn2011outcome}; requiring more complex randomisation systems which in turn may impact things such as drug supply and manufacture. When the main advantage pursued is patient benefit, this may compromise other characteristics of the design; for example in a two-arm trial RAR designs will require larger sample sizes than a traditional fixed design  with equal sample sizes in both arms; methods to account for such compromise have been proposed \cite{villar2018response,viele2019comparison}.\\

Choosing an approach from the existing variants of RAR can be challenging; in most cases, balancing the need for maximising statistical power and randomly assigning patients in an ethical manner is required. Most RAR methods require the availability of a reliable short-term outcome (although the exact form of the data may vary \cite{villar2018response,smith2018bayesian}); however, most of them can result in bias; therefore requiring the use of extra correction methods if collected data is used for estimation purposes \cite{bowden2017unbiased}. Another statistical concern is control of the type~I and type~II error rates. As discussed above this is possible but requires intensive simulations or the use of specific theoretical results \cite{hu2006theory,hu2003optimality,berry2010adaptive,simon2011using}; this creates an additional burden at the design stage, requiring additional time and support. 

\subsubsection{MCP-Mod}

The adaptive designs described in this section so far make no assumption about the relationship between the treatment arms, i.e. it is not required that one can order the arms with respect to the increasing efficacy.\\

In Phase~II dose-ranging studies, patients are typically randomised to either one of a number of doses and possibly a placebo. The objective of the trial is to find which dose (if any) should be tested in Phase~III. The target dose is often the minimum effective dose, the smallest dose giving a particular clinically relevant effect.\\

A traditional approach to find the target dose is based on pairwise comparisons. This approach, however, uses the information from the corresponding doses only and typically results in larger sample sizes required in the trial~\cite{bretz2017book}. As an alternative MCP-Mod \cite{bretz2005combining,pinheiro2014model} employs a dose-response model and allows for interpolation between the doses. \\

\begin{figure}[!ht]
\makebox[\textwidth][c]{\includegraphics[scale=.8]{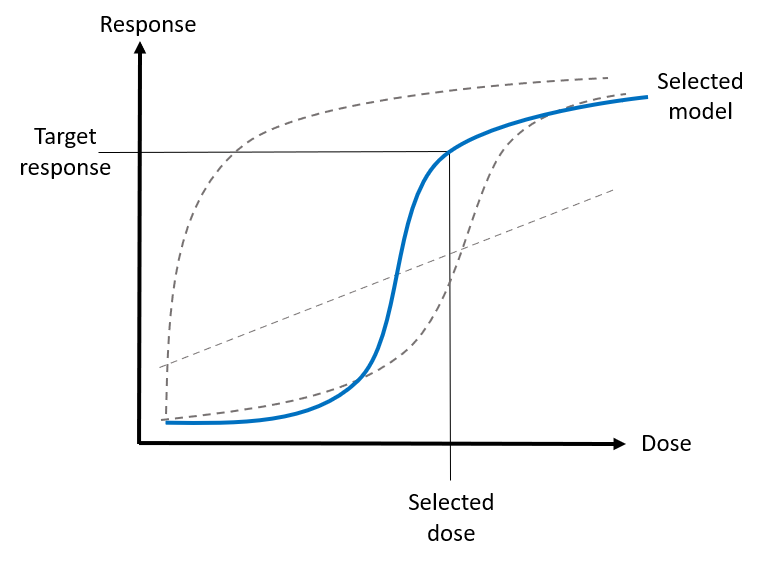}}
\caption{Model fitting in MCP-Mod.}
\label{fig:MCP}
\end{figure}

MCP-Mod is a two-stage method that combines Multiple Comparison Procedures (of dose levels) and MODeling approaches. At the planning stage, the set of possible models for the relationship between dose and response are defined, such as those shown in Figure~\ref{fig:MCP}. The inclusion of several models addresses the issue of some of the models being potentially mis-specified. At the trial stage, the MCP step checks whether there is any dose-response signal. This is done through hypothesis tests for each model while adjusting for the fact that there are multiple candidate models. This ensures control of the probability of making incorrect claims of a dose-response signal (a type $\mathrm{I}$ error). If no models are found to be significant, it is concluded that the dose-response signal cannot be detected given the observed data. \\

Once the dose-response signal is established, a single model is selected, or if multiple models are selected, an average over these models is made. The selection of models can be based either on tests performed at the MCP step or on some other measures such as the Akaike Information Criterion \cite{sakamoto1986akaike} (a measure of fit of statistical models). The selected model is then used to determine the dose to be tested in the subsequent trial. We refer the reader to the works focusing on the step-by-step application of MCP-Mod in practice \cite{bornkamp2009mcpmod, bretz2017book}. \\

\underline{Example:} Verrier et al.~\cite{verrier2014dose} described the application of MCP-Mod in a placebo controlled parallel group study undertaken in hypercholesterolaemic patients, which evaluated the change in low-density lipoprotein cholesterol (LDLC, mg/dL) following 12-weeks treatment as the primary endpoint. Three active doses were studied: 50, 100, and 150 mg, and nearly 30 patients per treatment group were recruited. The objective of the trial was two-fold: (i) to demonstrate the dose-effect of the compound, and (ii) to select the dose providing at least a 50\% decrease in LDLC.\\

The set of pre-specified candidate models was composed of linear, logistic (four possible pairs of parameters obtained from the guess that 50\% of the maximum effect occured at 50, 75, 100, or 125 mg, respectively, defined four possible models), and quadratic (corresponding to the maximum effect at 125 mg) models. The model selected by the hypothesis tests was a logistic model and the estimated target dose was 76.7 mg. Alternatively using an information criterion approach  the selected model was the quadratic model and the estimated target dose was 78.2 mg. To check the robustness of the results, model-averaging was also used and resulted in nearly the same estimated target dose. This analysis informed the selection of the dose for Phase~III trials for which the dose of 75 mg was chosen. This choice of dose was not one of those three active doses directly studied but could be selected due to using a model-based approach.\\

\underline{Advantages:} MCP-Mod allows for a more efficient use of data. Since its proposal, MCP-Mod was thoroughly studied, and many practical recommendations are available~\cite{bretz2017book}. MCP-Mod has been successfully applied in a number of trials \cite{bornkamp2013ema} as primary or secondary analysis, and the European Medicines Agency issued a qualification opinion of MCP-Mod~\cite{ema2014} concluding that MCP-Mod uses available data better than the traditional pairwise comparisons; and the FDA also designated the method as fit for purpose \cite{FDA2020}. There is software that implements the methodology, e.g. an \texttt{R}-package, \texttt{DoseFinding}~\cite{dosefindingpackage} and \texttt{PROC MCPMOD} in SAS. \\

\underline{Disadvantages:} The method can be sensitive to the model assumptions, specifically, to the initial guess of the model parameters~\cite{fda2015}. This can result in significantly lower power if the dose response relationship is not well approximated by one of the pre-specified candidate models. The number of doses to be included should inform the choice of the candidate models. Finally, the application of MCP-Mod should be approached with much care or avoided at all when the treatment regimens consist of various drugs, schedules and doses.

\subsubsection{Summary}
The methods presented in this section are suitable for selecting a treatment or dose for further study or, if constructed correctly, even allow for formal testing in a confirmatory setting. With RAR we see the goal of focusing on the more effective treatments while studying several was thought of as an important topic almost 90 years ago; however, it is only in the last 30 years or so that this topic has gained traction as a more active area of methodological research. These more modern methods will need to be sufficiently well understood before we start to see their use in clinical practice \cite{angus2019adaptive}.\\

Each of the methods discussed use adaptation rules to make efficient comparisons of several treatments or doses, filling a niche of application. There is a common advantage of reducing the number of patients (or expected number of patients) required to achieve the same strength of evidence when compared with fixed sampling alternatives. The key challenge is making informed decisions given the uncertainty about how the trial will develop before patient information is observed, whether that be how many patients will be required overall or how many patients will be allocated to each treatment. \\

For RAR, MAMS and DTL designs the trial is able to focus on those treatments that are demonstrating effectiveness. For trial participants and from an ethical perspective these methods are appealing as they increase the chance of receiving a treatment that is more likely to be effective. Further to this the model-based approach of MCP-Mod increases understanding of the relationship between dose and response in order to better allocate patients based on current trial data, which in turn allows greater confidence about the choice of dose for further study. \\

\subsection{Which patients will benefit?}

Late into the development cycle, for example in Phase~III of drug development, we wish to confirm the treatment works by demonstrating efficacy. An important aspect of this is to ensure the right patients receive the treatment (i.e. those who will gain a meaningful benefit from the treatment). Here, we focus on randomised clinical trials that use clinically relevant biomarkers to identify patients who may be sensitive to a treatment and therefore likely to respond. This is usually the case when several new therapies and corresponding predictive biomarkers are available for testing within a disease setting. 

\subsubsection{Covariate-Adjusted Response Adaptive (CARA)}

A form of RAR (see Section~\ref{sec:RAR}) that may be useful for trials in the context of population selection is Covariate Adjusted Response Adaptive (CARA) where randomisation probabilities are aligned to not only the observed efficacy of arms but also the patient's observed biomarker information. CARA skews allocation probabilities towards the best performing arms according to a patient's set of characteristics. Such changes in randomisation probabilities based upon available biomarker information are one of the most common adaptations in biomarker adaptive designs \cite{antoniou2016biomarker}.\\

CARA procedures are sometimes (incorrectly) referred to as minimisation procedures or dynamic allocation; some of the methods referred to with these names are essentially very different in their goals and nature. For example, some CARA procedures have been proposed altering the randomisation (similar to RAR) \cite{pocock1975sequential} while other methods do not do so in a randomised fashion determining allocation probabilities based solely on covariates \cite{taves1974minimization,altman2005treatment,scott2002method}. Some CARA procedures are designed to minimise imbalances on important covariates only \cite{pocock1975sequential,taves1974minimization} (rather than giving the best treatment for a given patient) while other methods have an efficiency goal, being designed to minimise the variance of the treatment effect in the presence of covariates (i.e. to maximise statistical power) \cite{atkinson1982optimum}. Finally, some CARA rules will aim to assign the largest number of patients to the best treatment while accounting for patients’ differences in biomarkers (as long as they are relevant to determining their best treatment) \cite{rosenberger2001covariate}.\\

\underline{Example:}  The BATTLE \cite{kim2011battle} trial is an example of a prospective, biopsy-mandated, biomarker-based, adaptively randomised \cite{zhou2008bayesian} study conducted in 255 pre-treated lung cancer patients. Following an initial equal randomisation period for the first 97 patients, 158 patients were adaptively randomised to four arms: erlotinib, vandetanib, erlotinib plus bexarotene, or sorafenib, based on relevant molecular biomarkers. The allocation probabilities were adapted based on the observed signal of interaction between the arms and biomarkers using a CARA procedure. Results include a $46\%$ 8-week disease control rate (primary endpoint) and evidence of an impressive benefit from sorafenib among mutant-KRAS patients.\\

\underline{Advantages/Disadvantages:} Because they fall within the RAR procedures, advantages and disadvantages of CARA designs are almost the same as with RAR. The main advantage of incorporating CARA rules into biomarker clinical trials is that they allow the flexibility of introducing balance, efficiency or ethical goals according to what might be more relevant for a particular trial.\\

\subsubsection{Population Enrichment}

\begin{figure}[!ht]
\makebox[\textwidth][c]{\includegraphics[scale=.8]{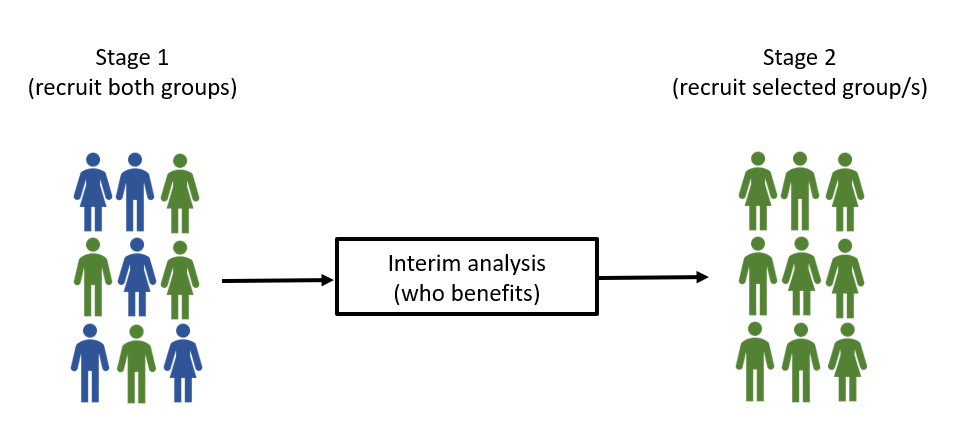}}
\caption{An adaptive enrichment design examining 2 subgroups.}
\label{fig:AE}
\end{figure}

Population Enrichment designs are useful when two or more biomarker-defined subgroups are known before the trial commences. With uncertainty about which populations should be recruited for the trial they ensure patients are recruited from appropriate populations using interim analyses to select which sub-populations to recruit from for the remainder of the trial. Figure~\ref{fig:AE} shows an example of how an adaptive enrichment design can progress. In a non-adaptive trial in the equivalent setting, the study team must make this sub-population selection before the trial begins without the benefit of trial observations. Population Enrichment designs are typically planned with one of two goals in mind: target the sub-population where patients receive the greatest benefit; or alternatively stop recruiting from sub-populations where the treatment may not provide a benefit.\\

Fully flexible hypothesis testing methods \cite{bretz2006confirmatory,schmidli2006confirmatory,marcus1976closed} may be constructed to preserve the statistical integrity of the trial while allowing freedom in the way decisions are made about sub-populations. The decision making methodology is a key design element in an adaptive population enrichment trial to which there are several approaches, both Bayesian \cite{brannath2009confirmatory,burnett2017,ondra2017optimized} and classical techniques such as setting thresholds for the observed effects being applied \cite{gotte2015improving,magnusson2013group}.\\

\underline{Example:} TAPPAS \cite{jones2017tappas,mehta2019adaptive} is a trial of TRC105 (an antibody) and pazopanib versus pazopanib alone in patients with advanced angiosarcoma. The study identified two sub-groups, those with cutaneous advanced angiosarcoma and those with non-cutaneous advanced angiosarcoma. Since there was some indication of greater tumour sensitivity to TRC105 in the cutaneous sub-group an adaptive enrichment design was used with the option to restrict enrolment to the cutaneous subgroup based on a interim analysis. The primary endpoint for this study was progression free survival, with an initial sample size of 124 patients to be followed until 95 events (progression or death) have been observed.\\

At the interim analysis the data monitoring committee were able to recommend one of three pre-planned actions: continue as planned  with  the  full  population (recruiting 124 patients followed until 95 events in total); continue  with  the  full  population and an increase in sample size and progression free survival events (recruiting 200 patients followed until 170 events in total); continue with only the cutaneous subgroup, thereby enriching the study population (recruiting 170 patients followed until 110 events in total).\\

The study concluded having recruited a total of 128 patients (close to the targeted recruitment of 124 with the trial continuing as planned in the full population), with 64 patients randomised to each the experimental treatment and the control. It was concluded that TRC105did not demonstrate activity when combined with pazopanib.\\

\underline{Advantages:} Population enrichment designs recruit fewer patients from subgroups that do not benefit from the treatment  allowing the trial to focus on those patients who are receiving a worthwhile benefit.  In scenarios where there is uncertainty about which subgroups benefit from the new treatment, the adaptive method can offer an improvement over non-adaptive alternatives.  This is because it is able to offer the benefit of increasing the sample size if one subgroup is preferable without sacrificing the opportunity to test the new treatment in all patients before the observation of any patients.  In addition, population enrichment offers a compromise between the possible fixed designs (for example recruiting only from one of the two sub-groups, or the full population). This compromise can be appealing in cases where there is disagreement between which (sub-)population(s) should be recruited from.\\

\underline{Disadvantages:} Computing decision rules that will provide good overall trial performance is often non-trivial \cite{burnett2017}, increasing the workload in setting up the trial.  Similarly, evaluating the expected performance of the trial is a non-trivial exercise, typically requiring some form of simulation. Possible changes to eligibility criteria after the interim analysis make planning patient recruitment challenging. \\

\subsubsection{Summary}

In this section we have discussed designs suitable for phase~II and phase~III of clinical development with the aim to test the effectiveness of a biomarker guided approach to treatment. This is well aligned with the shift across healthcare towards personalised medicine, ensuring patients receive treatments appropriate to them. Despite active research in this area, these methods appear to have the lowest uptake at the time of writing, although with the drive towards personalised healthcare they are likely to become increasingly relevant.\\

The suitability of each design depends on whether predictive biomarkers have been defined or are yet to be fully developed and tested. The methods presented allow formal testing for efficacy of an experimental treatment in biomarker defined sub-populations. Of note for methods that assume there is a pre-defined biomarker, the methods assume these biomarker groups are well defined, with little work on how to properly account for this should this assumption be violated \cite{wan2019subgroup}. Beyond being structured to allow formal analysis of sub-groups there is also a large ethical benefit to these designs; exposing as few patients as possible to treatments from which they may not receive a benefit.\\

Should no such pre-defined biomarkers be available one might consider adaptive signature designs \cite{freidlin2005adaptive,bhattacharyya2019adaptive}, also referred to as biomarker adaptive designs \cite{chen2014biomarker,antoniou2016biomarker,wason2014adaptive}. They aim to identify and use predictive biomarkers \cite{chen2014biomarker} during the trial. They help to improve the chances of identifying patients who will benefit from the treatment, whilst still providing accurate treatment effect estimates; this is done while maximising the use of the data used in the construction of the biomaker itself. However, the identification of predictive biomarkers in itself is not a trivial task and adds a large amount of complexity when working in the setting of a confirmatory clinical trial. With the sub-groups undefined, this creates further uncertainty adding a large administrative burden to implementing such a design.\\

An extension beyond the methods presented is the wider scope of umbrella \cite{park2019systematic} and basket trials \cite{cunanan2017efficient}, which make use of more detailed biomarker information. Broadly speaking: a basket trial examines a single experimental  treatment in multiple sub-types of a single biomarker, while an umbrella trial may consider mutiple biomarkers each of which may require a different treatment.\\

\subsection{Does the treatment work?}

In this section, we consider treatments late in the development cycle, typically phase~III (although viable in phase~II also), where we wish to show conclusively that a novel treatment is an improvement over the current standard of care.  A conventional aim at this stage of development is to be efficient, both in terms of the time that is required to conduct the trial and the number of patients we must recruit; the methods that follow attempt to make the trial more efficient while also ensuring that the overall probability of a successful outcome is maintained or increased.  Methods aiming to be efficient in recruiting the correct number of patients have both an operational and ethical benefit; ensuring patients are not subject to an experimental treatment for a trial that is underpowered (i.e. has a low chance of detecting a meaningful effect) or when their contribution to the overall result is not required.\\

\subsubsection{Group sequential designs}
\label{sec:gs}
\begin{figure}[!ht]
\makebox[\textwidth][c]{\includegraphics[scale=.8]{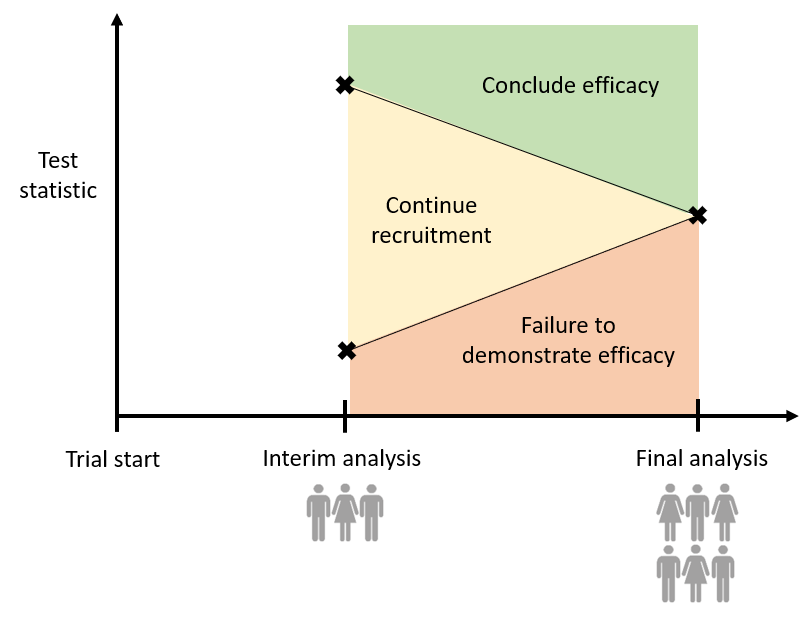}}
\caption{Demonstration of stopping boundaries in a 2 stage group-sequential design.}
\label{fig:GS}
\end{figure}

Group sequential designs are the most widespread \cite{hatfield2016adaptive} of the adaptive designs we consider in this paper.  A group sequential trial differs from a more traditional phase~III clinical trial in that one or more pre-planned interim analyses may be used to assess efficacy; if there is strong evidence the experimental treatment is superior to control, or indeed that both have the same effect, then the trial may be terminated early as depicted in Figure~\ref{fig:GS}.  Investigators must decide before the trial whether binding or non-binding futility boundaries are to be used, that is whether the early stopping for futility must be strictly adhered to, or whether they can be overruled without risking an inflation of the type~I error rate. This choice affects on the overall definition of the early stopping thresholds (both for superiority and for futility). \\

To enable early stopping of the trial while maintaining control of the type~I error rate as required in a confirmatory setting, group sequential trials use pre-defined stopping rules.  At each interim analysis the current data for the experimental and control arms are compared to construct a test statistic. If the test statistic is sufficiently high/low the trial is stopped early for efficacy/lack of a demonstrated benefit of the treatment; if neither criterion is met, the trial continues to the next interim or final analysis.\\

Several approaches to the definition of stopping rules for group sequential designs have been proposed \cite{pocock1977group,o1979multiple,gordon1983discrete,whitehead1997design}. Jennison \& Turnbull \cite{jennison1999group} provide a comprehensive guide to group sequential designs and their application. Whitehead (2011) \cite{whitehead2011group} describes the implementation of group sequential methods using SAS, in addition the gsDesign \cite{anderson2009gsdesign} and optGS \cite{wason2013optgs} packages in R are open source alternatives allowing the construction and analysis of group sequential designs.\\

\underline{Example:} The INTERCEPT trial \cite{boden2000diltiazem,of1995design} was a randomised, double blind trial in patients  with  acute myocardial  infarction. The trial used a group sequential design to achieve $80\%$ power to detect a $33\%$ between-group difference in cumulative first event rate of cardiac death, non-fatal reinfarction, or refractory ischaemia. Interim  analyses  were  conducted  after  enrolment  of  300 patients,  and  at  intervals of roughly 300 patients  thereafter. The stopping boundaries were constructed using a double triangular test \cite{whitehead1997design}.\\

Recruitment was stopped after the third interim analysis. In total the trial recruited  874  patients  with  acute myocardial  infarction,  but  without  congestive  heart  failure, who  first  received  thrombolytic  agents. 430 patients were randomised to 300 mg oral diltiazem once daily, and 444 patients were randomised to placebo initiated within  36–96  hours  of  infarct  onset,  and  given  for  up  to 6 months. It was concluded that the  pre-specified  $33\%$  between-group difference  for  the  primary  endpoint  would  probably  not  be achieved.\\

\underline{Advantages:} Group sequential designs offer a reduction in the expected sample size of the trial  compared to a traditional trial design with a fixed sample size. This is because there is a high probability that the trial will stop early due to strong evidence that the treatment either is or is not effective. The reduction in expected sample size compared to a fixed design, is typically around $15\%$ for O'Brien-Fleming \cite{o1979multiple} boundaries \cite{guidance2018adaptive}. Group sequential designs are the optimal choice in terms of minimising the expected sample size \cite{jennison1999group}.  The implication is that any other design aiming to reduce the expected sample size can only perform as well as, but never better than, the group sequential option.  In addition, with group sequential methods being widespread \cite{hatfield2016adaptive} their implementation is likely to be simpler than more novel adaptive designs, and also more familiar to regulators and ethics committees. \\

\underline{Disadvantages:} Taking multiple looks at the data to provide the opportunity for early stopping requires careful definition of stopping boundaries, and there is a cost to this adjustment. While the expectation is for the design to reduce the sample size by stopping early, it is possible that for a given trial the group sequential design will recruit more patients than a non-adaptive method would have (when no early stopping criterion is met), usually by approximately $10\%$ \cite{guidance2018adaptive}. Due to uncertainty about the length of the trial and number of patients before the trial commences, each stage of the trial may or may not be the last, making logistics and planning more complex than for a traditional fixed sample design.\\

\subsubsection{Sample size re-assessment}
\label{sec:ssr}
\begin{figure}[!ht]
\makebox[\textwidth][c]{\includegraphics[scale=.8]{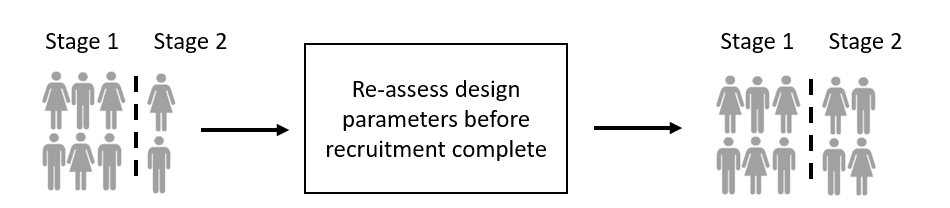}}
\caption{Sample size re-assessment re-assessing key parameters using an interim analysis to update the required sample size.}
\label{fig:SSR}
\end{figure}

Sample size re-assessment (also known as sample size re-estimation or sample size re-calculation) seeks to ensure the use of an appropriate sample size for the trial despite uncertainty about key parameters (such as the variance in the observations) during the designing of the trial by re-assessing the required sample size during the trial as shown in Figure~\ref{fig:SSR}.  Typically, the re-estimation of the sample size focuses on the estimation of design parameters at an interim analysis to re-calculate the sample size in order to achieve, for example, a desired conditional power \cite{proschan2009sample,friede2006sample,chuang2006sample} (the probability of rejecting the null hypothesis given the currently available data). Usually, this is done with the option to increase but not the option to decrease the sample size. In practice there is often a limit on what maximum sample size is possible, as large increases to the sample size in this way can be inefficient \cite{wang2012paradigms}; in practice if the estimated sample size exceeds the pre-set maximum then recruitment is usually stopped (a form of stopping for futility). \\

Sample size re-assessment may be either blinded or unblinded \cite{pritchett2015sample}, while maintaining the statistical integrity of the trial. In unblinded sample size re-estimation interim analyses are conducted based on unblinded trial data; that is, the statistician performing the interim analysis will know which participants are in which trial arm. This is similar to group sequential designs although the aim here is not to stop the trial early. Updated estimates of parameters related to the sample size are used to re-assess the required sample size for the remainder of the trial \cite{spiegelhalter2004bayesian,lachin2005review,broglio2014not}. Conversely, blinded sample size re-estimation only makes use of the combined blinded trial data from all treatment arms when making decisions at the interim analyses \cite{friede2006sample,proschan2009sample,zucker1999internal,kieser2003simple}. Alterations to the sample size in this way have a minimal impact on the type~I error rate, even without formal adjustment \cite{kieser2003simple}. The suitability of blinded or unblinded re-assessment will depend on the parameters that require re-estimation.\\

\underline{Example:} Hade et al. \cite{hade2019follow} discuss a sample size re-assessment in a randomised trial in breast cancer where the primary outcome is disease-free survival. Women were randomised either to: immediate surgery in the next 1--6 days, which was expected to be in the follicular phase of the menstrual cycle; or to scheduled surgery during the next mid-luteal phase of the menstrual cycle. Based on primary analysis by log-rank test with a target of $80\%$ power, with $5\%$ two-sided type~I error rate, to detect a Hazard Ratio (HR) of 0.58 in favour of scheduled surgery, this required 113 events. With accrual time of 2 years and 4 additional years of follow up and a $2-3\%$ loss to follow up, the initial study design planned to randomise 340 women. \\

The HR used for the proposed sample size was felt to be optimistic based on more current information and sources external to the trial. Based on a blinded sample size re-assessment using the available data, the sample size was increased by 170 patients, to a total of 510 randomised patients (with a required 175 events) in order to target a revised HR of 0.65.\\

\underline{Advantages:} The main aim of sample size re-assessment is to ensure that the trial recruits an appropriate number of patients, particularly when there is high uncertainty about the design parameters before the trial begins whilst ensuring the integrity of the trial is maintained.  Sample size re-assessment designs are not as complex as many other adaptive methods, which means the trial may be planned and started more quickly.  The fact that both unblinded and blinded methods are available means that sample size re-assessment can be applied in many different trial settings. There is an upward trend in the use of sample size re-assessment in clinical practice, and as these designs become more widespread it will become increasingly easy to put them forward.\\

\underline{Disadvantages:} This is a method with relatively few drawbacks. There is a small additional burden for statistical input at the interim analysis to properly estimate the required sample size for the remainder of the trial, which requires appropriate expertise in the case of both blinded and unblinded interim analyses. Most of the practical issues that a sample size re-assessment design may bring (for example time constraints or securing sufficing funding in advance) are similar to those faced when using other adaptive designs.\\

\subsubsection{Summary}

Methods to confirm a treatment works are the most widespread of the adaptive designs we have considered \cite{hatfield2016adaptive}, to the extent group sequential trials may even be considered a standard approach.  The group sequential framework forms a foundation for many other adaptive methods due to its preservation of the integrity of the trial results (for example, we discussed how MAMS designs may be constructed to control the type~I error rate, one method for this is to construct group sequential-type tests).\\ 

\section{Conclusions}

Research into adaptive designs has become more prevalent across all stages of the clinical development process, although this increase is not necessarily reflected by their uptake in clinical practice. The suitability of an adaptive method depends largely on the clinical question being addressed. We have presented four key clinical questions for which corresponding adaptive designs may be of use across a wide range of disease areas, study settings and endpoints. For each possible design, there are advantages and disadvantages but in general there are some key themes: there is an increase in efficiency of the design in terms of the expected number of patients or a clear benefit in understanding the question of scientific interest; there are clear ethical advantages to ensuring the right patients are given the best available treatment whenever possible; and the key disadvantage is the additional burden, both in the planning the trial and the interim analyses. Importantly, while the adaptive methods can be highly effective when used in the correct scenario it is not always the case an adaptive method is the best choice \cite{wason2019keep} and so careful consideration must be taken as to if it is the correct choice.\\

Regulatory bodies are increasingly recognising the desire for the use of adaptive designs and accepting their use  although it is recommended regulators be engaged early in the process whenever using any novel methodology \cite{guidance2019adaptive,committee2007reflection}.  Funding bodies are also increasingly comfortable with the use of adaptive designs,  the TAILoR trial discussed in Section~\ref{sec:MAMS} appears as a case study on the National Institute for Health Research website \cite{NIHR2019}.  Additionally new reporting guidance for adaptive designs \cite{dimairo2019adaptive} have recently been published to facilitate uptake further. \\

We have not been exhaustive in our discussion of adaptive designs, focusing on the key designs to answer the most common questions of clinical interest. Seamless designs \cite{bretz2006confirmatory,schmidli2006confirmatory,jennison2006confirmatory}, which we have not discussed in detail, use similar adaptive design methodology to combine phases of clinical development: designs may be inferentially seamless, where data from the earlier stage are incorporated into the overall trial results; operationally seamless, avoiding any break in recruitment between the stages of the development process but excluding data from earlier stages from the final analysis of the latter; or both.  There are many motivations for conducting seamless designs \cite{cuffe2014seamless}, making it an active area of research \cite{graham2019comparison}.\\

For many years one major obstacle to the use of adaptive designs in practice has been the lack of suitable software to aid both the design and conduct of trials. This issue is increasingly being tackled by those researching the methods, with many open source packages available for the design and analysis of adaptive methods some of which have been cited in this work. For example rpact \cite{wassmer2019rpact} is an R package that assists in the design and analysis of confirmatory clinical trials. In addition, there is a steep learning curve to the implementation of such designs; training courses are becoming increasingly available to address this.\\

From a methodology standpoint there are some further issues that go beyond the level of detail we have discussed that should be considered when proposing an adaptive design, for example potential information leakage or the introduction of bias \cite{sanchez2014practical,chow2012independence}. The Practical Adaptive and Novel Designs and Analysis (PANDA) toolkit \cite{PANDA} is under development at the time of writing and will be an online resource that addresses and explains broader issues in the use of adaptive designs.\\

Despite the challenges in the design and analysis of an adaptive trial we believe that under the right circumstance the benefits introduced by the increased flexibility clearly outweigh these issues.\\

\begin{backmatter}

\section*{Competing interests}
  The authors declare that they have no competing interests.

\section*{Acknowledgements}
TB was supported by the MRC Network of hubs for Trials Methodology HTMR Award MR/L004933/1. SSV is supported by UK Medical Research Council (grant number: MC\_UU\_00002/3). GMW is supported by Cancer Research UK. This report is independent research arising in part from Prof Jaki’s Senior Research Fellowship (NIHR-SRF-2015-08-001) supported by the National Institute for Health Research. The views expressed in this publication are those of the authors and not necessarily those of the NHS, the National Institute for Health Research or the Department of Health and Social Care (DHCS). TJ is also supported by	UK Medical Research Council (grant number: MC\_UU\_0002/14).

\bibliographystyle{bmc-mathphys}
\bibliography{bmc_adaptivedesigns}

\section*{Tables}

\begin{table}[p]
\makebox[\linewidth]{
\begin{tabular}{p{0.1\textheight} | p{0.1\textwidth} | p{0.1\textwidth} | p{0.35\textwidth} | p{0.35\textwidth}}
Method & a.k.a & Phase of development & Definition & Benefits \\ \hline

Adaptive treatment switching &  & $\mathrm{II/III}$ & allow trial participants to switch from allocated treatment to an alternative & more trial participants receive preferred  treatment\\ \hline

Bayesian adaptive &  & $\mathrm{I/II/III}$ & Bayesian methodology may be incorporated into many other designs in the analysis and/or the interim decision making & lower sample size due to utilisation of prior information\\ \hline

Biomarker adaptive & Adaptive signature & $\mathrm{II/III}$  & identify and utilise biomarker information to modify trial in progress to target population & target the correct patient population\\ \hline

Covariate adjusted response adaptive & CARA & $\mathrm{II/III}$ & shift allocation ratio towards promising treatment(s) using covariate information & more trial participants receive effective treatment \\\hline

Continual reassessment method & CRM & $\mathrm{I}$ & dose escalation design for finding the maximum tolerated dose (MTD) & more accurate and precise estimation of the MTD than with 3+3 designs, more patients treated at or close to the MTD\\ \hline

Drop-the-loser & DTL & $\mathrm{II/III}$ & drop inferior treatment arms (control group typically retained) &  fewer trial participants assigned to less effective treatments\\ \hline

Escalation with overdose control & EWOC & $\mathrm{I}$ & dose escalation design to find MTD using an allocation criterion to avoid overdosing &  more accurate and precise estimation of the MTD than with 3+3 designs, avoiding undesirable overdosing of patients\\\hline

Group sequential design &  & $\mathrm{II}/\mathrm{III}$ & early stopping for futility or efficacy & reduction in the expected sample size, typically allowing for faster trials requiring fewer patients (for a small increase in the possible maximum sample size) \\ \hline

Multi-arm multi-stage trial & MAMS & $\mathrm{II/III}$ & compare multiple treatments to a common control, allow for early stopping for efficacy or futility & common control requires fewer patients than conducting separate trials, early stopping for efficacy or futility \\ \hline

MCP-Mod & & $\mathrm{II}$ & combination of multiple comparisons and modeling approaches to establish dose response model & efficient use of available data vs pairwise comparisons \\\hline

Population enrichment & Adaptive enrichment & $\mathrm{II/III}$ & allow for selection of target population during the trial based on pre-defined patient populations & target the correct patient population \\ \hline

Response adaptive randomisation & Adaptive allocation, RAR & $\mathrm{II/III}$  & shift allocation ratio towards more promising treatment(s) & more trial participants receive effective treatment \\ \hline

Sample size re-assessment & Sample size re-estimation/re-calculation & $\mathrm{II/III}$ & mid-course adjustment of the sample size, in either a blinded or unblinded fashion & raise the probability of a successful trial\\ \hline

Seamless design & Portfolio decision making & $\mathrm{I/II/III}$ & merge trials from different phases of development e.g. phase~I/II or phase~II/III, can be inferentially and/or operationally seamless & (inferential) more efficient use of data from each phase of clinical development / (operationl) faster clinical development process and moving between stages
\end{tabular}}
\caption{Glossary of adaptive designs and descriptions of their typical applications}
\label{tab:glos}
\end{table}

\end{backmatter}
\end{document}